\documentclass[aps,prd,twocolumn,showpacs,superscriptaddress]{revtex4}
\usepackage{graphics}
\usepackage{psfig}
\usepackage{tabularx}

\begin{document}
\title{$SU(3)_c\otimes SU(4)_L\otimes U(1)_X$ without exotic electric 
charges}
\author{William A. Ponce} 
\affiliation{Instituto de F\'\i sica, Universidad de Antioquia,
A.A. 1226, Medell\'\i n, Colombia}
\author{Diego A. Guti\'errez}
\affiliation{Instituto de  F\'\i sica, 
Universidad de Antioquia,
A.A. 1226, Medell\'\i n, Colombia}
\author{Luis A. S\'anchez}
\affiliation{Escuela de F\'\i
sica, Universidad Nacional de Colombia,
A.A. 3840, Medell\'\i n, Colombia}


\begin{abstract}
We present an extension of the Standard Model to the local gauge group
$SU(3)_c\otimes SU(4)_L\otimes U(1)_X$ with a family non-universal
treatment and anomalies canceled among the three families in a nontrivial
fashion. The mass scales, the gauge boson masses, and the masses for the
spin $1/2$ particles in the model are analyzed. The neutral currents coupled to all neutral vector bosons in the model are studied, and
particular values of the parameters are used in order to simplify the
mixing between the three neutral currents present in the theory, mixing which is further constrained by experimental results from the CERN LEP, SLAC Linear Collider and atomic parity violation.
\end{abstract}

\maketitle

\section{\label{sec:intr}Introduction} 
In spite of the overwhelming phenomenological success of the standard
model (SM) based on the local gauge group $SU(3)_c\otimes SU(2)_L\otimes
U(1)_Y$, with $SU(2)_L\otimes U(1)_Y$ hidden and $SU(3)_c$
confined \cite{sm}, it fails to explain several issues like hierarchical
fermion masses and mixing angles, charge quantization, strong $CP$
violation, replication of families and neutrino oscillations among others.  
For example in the weak basis, before symmetry is broken, the three
families in the SM are identical to each other; when symmetry breaking
takes place, the fermions get masses according to their experimental
values and the three families acquire a strong hierarchy. However in the
SM there is no mechanism for explaining the origin of families or
the fermion mass spectrum.

These drawbacks of the SM have led to a strong belief that the model is
still incomplete and that it must be regarded as a low-energy effective
field theory originating from a more fundamental one. That belief lies on
strong conceptual indications for physics beyond the SM which have
produced a variety of theoretically well motivated extensions of the
model: left-right symmetry, grand unification, supersymmetry, superstring
inspired extensions, etc. \cite{ext}.

At present the only experimental fact that points toward a beyond the SM 
structure lies in the neutrino sector, and even there the results are not 
final yet. So a reasonable approach is to depart from the SM as 
little as possible, allowing some room for neutrino 
oscillations \cite{neutrinos}.

$SU(4)_L\otimes U(1)_X$ as a flavor group has been considered before in
the literature \cite{su4, little}, and, among its best features, provides 
an alternative to the problem of the number $N_f$ of families, in the sense
that anomaly cancellation is achieved when $N_f=N_c=3, \; N_c$ being the
number of colors of $SU(3)_c$ (also known as QCD). In addition, this gauge
structure has been used recently in order to implement the so-called
little Higgs mechanism \cite{little}.

In this paper an analysis of the $SU(3)_c\otimes SU(4)_L\otimes U(1)_X$
local gauge theory (hereafter the 3-4-1 theory) shows that, by restricting
the fermion field representations to particles without exotic electric
charges and by paying due attention to anomaly cancellation, a few
different models are obtained, while by relaxing the condition of the
nonexistence of exotic electric charges, an infinite number of models can
be generated.

This paper is organized as follows. In the next section we introduce the
model based on the local gauge group $SU(3)_c\otimes SU(4)_L\otimes
U(1)_X$ which we are going to study. In Sec.~\ref{sec:sec3} we describe
the scalar sector needed to break the symmetry and to produce masses to
the fermion fields in the model. In Sec.~\ref{sec:sec4} we study the gauge
boson sector paying special attention to the neutral currents present in
the model and their mixing. In Sec.~\ref{sec:sec5} we analyze the fermion
mass spectrum. In Sec.~\ref{sec:sec6} we use experimental results in order
to constrain the mixing angle between two of the neutral currents and the
mass scale of the new neutral gauge bosons. In the last section we
summarize the model and state our conclusions. At the end an Appendix is
presented in which we make a systematic analysis of the 3-4-1 symmetry and
obtain general conditions to have anomaly free models without exotic
electric charges.
  

\section{\label{sec:sec2}The fermion content of the model} 
In what follows we assume that the electroweak gauge group is
$SU(4)_L\otimes U(1)_X$ which contains $SU(2)_L\otimes U(1)_Y$ as a
subgroup, with a non-universal hypercharge $X$ in the quark sector, which
in turn implies anomaly cancellation among the families in a non-trivial
fashion. We also assume that the left-handed quarks (color triplets)  and
left-handed leptons (color singlets) transform either under the 4 or
$\bar{4}$ fundamental representations of $SU(4)_L$, and that as in the SM, 
$SU(3)_c$ is vectorlike.

With the former assumptions we look for the simplest structure in such a 
way that, not only it does not contain fields with exotic electric 
charges, but also that charged exotic leptons are absent from the 
anomaly-free spectrum. According to the Appendix there is only one model 
(Model {\bf A}) satisfying all those constraints, for which the electric 
charge operator is given by 
$Q=T_{3L}+T_{8L}/\sqrt{3}+ T_{15L}/\sqrt{6}+ XI_4,$
with the following fermion structure:

\[\begin{array}{ccccc}\hline\hline
Q_{aL}=\left(\begin{array}{c}u_a\\d_a\\D_a\\D'_a \end{array}\right)_L &
u^c_{aL} & d^c_{aL}& D^c_{aL}& D^{'c}_{aL} \\ \hline [3,4,-\frac{1}{12}] &
[\bar{3},1,-{2\over 3}] & [\bar{3},1,{1\over 3}] & [\bar{3},1,{1\over 3}]&
[\bar{3},1,{1\over 3}] \\ \hline\hline \end{array} \]

\[\begin{array}{ccccc}\hline\hline
Q_{1L}=\left(\begin{array}{c}d_1\\u_1\\U_1\\U^\prime_1 
\end{array}\right)_L & 
d^c_{1L} & u^c_{1L}&
U^c_{1L}& U^{'c}_{1L} \\ \hline [3,\bar{4},{5\over 12}] & 
[\bar{3},1,{1\over 3}] & [\bar{3},1,-{2\over 3}]
& [\bar{3},1,-{2\over 3}]& [\bar{3},1,-{2\over 3}] \\ \hline\hline 
\end{array} \]

\[\begin{array}{cc}\hline\hline
L_{\alpha, L}=\left(\begin{array}{c} e^-_\alpha\\ \nu_{e \alpha}\\ 
N^0_\alpha \\ N^{'0}_\alpha \end{array}\right)_L & 
e^+_\alpha \\ \hline
[1,\bar{4},-{1\over 4}] & [1,1,1] \\ \hline\hline
\end{array} \]

\noindent
where $a=2,3$ and $\alpha = 1,2,3$ are two and three family indexes,
respectively. The numbers in parentheses refer to the $[SU(3)_C, SU(4)_L,
U(1)_X]$ quantum numbers, respectively. Notice that if needed, the lepton
structure of the model can be augmented with an undetermined number of
neutral Weyl singlet states $N^{0}_{L,b} \sim [1,1,0]$, $b=1,2,...,$ 
without violating our assumptions, neither the anomaly constraint relations, because singlets with no $X$ charges are as good as not being present as far as anomaly cancellation is concerned.


\section{\label{sec:sec3}The scalar sector} 
Our aim is to break the symmetry following the pattern

\begin{eqnarray}\nonumber
SU(3)_c\otimes SU(4)_L\otimes & U(1)_X & \\* \nonumber  
& \rightarrow & SU(3)_c\otimes SU(3)_L\otimes
U(1)_X \\* \nonumber 
& \rightarrow & SU(3)_c\otimes SU(2)_L\otimes U(1)_Y \\* \nonumber 
& \rightarrow & SU(3)_c\otimes U(1)_Q, \nonumber
\end{eqnarray}
where $SU(3)_c\otimes SU(3)_L\otimes U(1)_X$ refers to the so-called
3-3-1 structure introduced in Ref.~\cite{331}. At the same time we want to 
give masses to the fermion fields in the model.  With 
this in mind we introduce the following three Higgs scalars: 
$\phi_1[1,4,-3/4]$ with a vacuum expectation value (VEV) aligned in the 
direction $\langle\phi_1\rangle=(v,0,0,0)^T$; $\phi_2[1,\bar{4},-1/4]$ 
with a VEV aligned as $\langle\phi_2\rangle=(0,0,V,0)^T$ and 
$\phi_3[1,\bar{4},-1/4]$
with a VEV aligned as $\langle\phi_3\rangle=(0,0,0,V^\prime)^T$, with the
hierarchy $V\sim V^\prime >> v\sim 174$ GeV (the electroweak breaking 
scale).


\section{\label{sec:sec4}The gauge boson sector}
In the model there are a total of 24 gauge bosons: One gauge field
$B^\mu$ associated with $U(1)_X$, the 8 gluon fields associated
with $SU(3)_c$ which remain massless after breaking the symmetry, and 
another 15 gauge fields associated with $SU(4)_L$ which we may write as 
\[{1\over 2}\lambda_\alpha A^\mu_\alpha={1\over \sqrt{2}}\left(
\begin{array}{cccc}D^\mu_1 & W^{+\mu} & K^{+\mu} & X^{+\mu}\\ W^{-\mu} &
D^\mu_2 &  K^{0\mu} &  X^{0\mu}\\
K^{-\mu} & \bar{K}^{0\mu} & D^\mu_3 & Y^{0\mu}\\
X^{-\mu} & \bar{X}^{0\mu} & \bar{Y}^{0\mu} & D^\mu_4 \end{array}\right), \]
where $D^\mu_1=A_3^\mu/\sqrt{2}+A_8^\mu/\sqrt{6}+A_{15}^\mu/\sqrt{12},\;
D^\mu_2=-A_3^\mu/\sqrt{2}+A_8^\mu/\sqrt{6}+A_{15}^\mu/\sqrt{12}$;
$D^\mu_3=-2A_8^\mu/\sqrt{6}+A_{15}^\mu/\sqrt{12}$, and 
$D^\mu_4=-3 A_{15}^\mu/\sqrt{12}$.

After breaking the symmetry with $\langle\phi_1\rangle +
\langle\phi_2\rangle+ \langle\phi_3\rangle$ and using for the covariant
derivative for 4-plets $iD^\mu=i\partial^\mu-g \lambda_\alpha
A^\mu_\alpha/2-g'XB^\mu$, where $g$ and $g^\prime$ are the $SU(4)_L$ and 
$U(1)_X$ gauge coupling constants respectively, we get the following mass 
terms for the charged gauge bosons: 
$M^2_{W^\pm}=g^2v^2/2$ as in the SM, $M^2_{K^\pm}=g^2(v^2+V^2)/2$, 
$M^2_{X^\pm}=g^2(v^2+V^{\prime 2})/2$,
$M^2_{K^0(\bar{K}^0)}=g^2V^2/2$, $M^2_{X^0(\bar{X}^0)}=g^2V^{\prime 2}/2$ and $M^2_{Y^0(\bar{Y}^0)}=g^2(V^2+V^{\prime 2})/2$. 
Since $W^\pm$ does not mix with $K^\pm$ or with $X^\pm$ we have that 
$v\approx 174$ GeV as in the SM.

For the four neutral gauge bosons we get mass terms of the form 

\begin{eqnarray*}
M&=&{g^2\over 2}\Big\{V^2 \left(\frac{g'B^\mu}{2g}
-\frac{2A_8^\mu}{\sqrt{3}}+\frac{A^\mu_{15}}{\sqrt{6}}\right)^2 \\
& & + V^{\prime 2}\left(\frac{g'B^\mu}{2g}-\frac{3 
A^\mu_{15}}{\sqrt{6}}\right)^2 \\
& &+v^2\left(A^\mu_3
+\frac{A_8^\mu}{\sqrt{3}}+\frac{A^\mu_{15}}{\sqrt{6}}
-\frac{3g'B^\mu}{2g}\right)^2\Big\}.
\end{eqnarray*}
\noindent 
$M$ is a $4 \times 4$ matrix with a zero eigenvalue
corresponding to the photon. Once the photon field has been identified, we remain with a $3 \times 3$ mass matrix for three neutral gauge bosons
$Z^\mu$, $Z^{'\mu}$ and $Z^{''\mu}$. Since we are interested now in 
the low energy phenomenology of our model, we can choose $V=V^\prime$ in order to simplify matters. For this particular case 
the field $Z''^\mu= A_8^\mu / \sqrt{3}-\sqrt{2/3}A_{15}^\mu$ decouples from the other two and acquires a squared mass $(g^2/2)V^2$.
By diagonalizing the remaining $2 \times 2$ mass matrix we get the other two physical neutral gauge bosons which are defined through the mixing angle $\theta$ between $Z_\mu,\; Z'_\mu$: 

\begin{eqnarray}\nonumber
Z_1^\mu&=&Z_\mu \cos\theta+Z'_\mu \sin\theta \; ,\\ \nonumber
Z_2^\mu&=&-Z_\mu \sin\theta+Z'_\mu \cos\theta, \end{eqnarray} 
where
\begin{equation} \label{tan} \tan(2\theta) = - \frac{2 \sqrt{2} C_W}
{\sqrt{1+2 \delta^2}\left[1+ \frac{2V^2}{v^2}C_W^4- \frac{2}{1+2
\delta^2}C^2_W \right]}, \end{equation}
with $\delta= g'/(2g)$.

\noindent  
The photon field $A^\mu$ and the fields $Z_\mu$ and $Z'_\mu$ are given by

\begin{eqnarray} \nonumber
A^\mu&=&S_W A_3^\mu \nonumber \\
& & + C_W\left[\frac{T_W}{\sqrt{3}}\left(A_8^\mu+
\frac{A_{15}^\mu}{\sqrt{2}}\right)+(1-T_W^2/2)^{1/2}B^\mu\right]\; , \nonumber \\  
Z^\mu&=& C_W A_3^\mu \nonumber \\
& & - S_W\left[\frac{T_W}{\sqrt{3}}\left(A_8^\mu+
\frac{A_{15}^\mu}{\sqrt{2}}\right)+(1-T_W^2/2)^{1/2}B^\mu\right] \; , \nonumber \\ \label{fzzp}
Z'^\mu&=&\sqrt{\frac{2}{3}}(1-T_W^2/2)^{1/2}\left(A_8^\mu+
\frac{A_{15}^\mu}{\sqrt{2}}\right)-\frac{T_W}{\sqrt{2}}B^\mu.
\end{eqnarray}
\noindent
$S_W=2 \delta /\sqrt{6 \delta^2 + 1}$ and $C_W$ are the sine and
cosine of the electroweak mixing angle respectively, and
$T_W=S_W/C_W$. We can also identify the $Y$ hypercharge associated
with the SM abelian gauge boson as
\begin{equation}\label{y}
Y^\mu=\left[\frac{T_W}{\sqrt{3}}\left(A_8^\mu+
\frac{A_{15}^\mu}{\sqrt{2}}\right)+(1-T_W^2/2)^{1/2}B^\mu\right].
\end{equation}

\subsection{\label{sec:sub4a}Charged currents}
After some algebra, the Hamiltonian for the charged currents can be 
written as 
\begin{widetext}
\begin{eqnarray}\nonumber
H^{CC}&=&{g\over \sqrt{2}}[ W^+_\mu \lbrace(\sum_{a=2}^3\bar{u}_{aL}\gamma^\mu 
d_{aL})- \bar{u}_{1L}\gamma^\mu d_{1L} - (\sum_{\alpha=1}^3
\bar{\nu}_{e\alpha L}\gamma^\mu e^-_{\alpha L})\rbrace \\ \nonumber & &
+K^+_\mu \lbrace(\sum_{a=2}^3\bar{u}_{aL}\gamma^\mu 
D_{aL})-\bar{U}_{1L}\gamma^\mu 
d_{1L} - (\sum_{\alpha=1}^3
\bar{N}^0_{\alpha L}\gamma^\mu e^-_{\alpha L})\rbrace \\ \nonumber & & 
+X^+_\mu\lbrace(\sum_{a=2}^3\bar{u}_{aL}\gamma^\mu 
D'_{aL})-\bar{U'}_{1L}\gamma^\mu 
d_{1L} - (\sum_{\alpha=1}^3
\bar{N'}^0_{\alpha L}\gamma^\mu e^-_{\alpha L})\rbrace \\ \nonumber & & 
+K^0_\mu\lbrace(\sum_{a=2}^3\bar{d}_{aL}\gamma^\mu 
D_{aL})-\bar{U}_{1L}\gamma^\mu 
u_{1L} - (\sum_{\alpha=1}^3
\bar{N}^0_{\alpha L}\gamma^\mu \nu_{e\alpha L})\rbrace \\ \nonumber & & 
+X^0_\mu\lbrace(\sum_{a=2}^3\bar{d}_{aL}\gamma^\mu 
D'_{aL})-\bar{U'}_{1L}\gamma^\mu 
u_{1L} - (\sum_{\alpha=1}^3
\bar{N'}^0_{\alpha L}\gamma^\mu \nu_{e\alpha L})\rbrace \\  & & 
+Y^0_\mu\lbrace(\sum_{a=2}^3\bar{D}_{aL}\gamma^\mu 
D'_{aL})-\bar{U'}_{1L}\gamma^\mu 
U'_{1L} - (\sum_{\alpha=1}^3
\bar{N'}^0_{\alpha L}\gamma^\mu N^0_{\alpha L})\rbrace ] + h.c.
\end{eqnarray}
\end{widetext}

\subsection{\label{sec:sub4b}Neutral currents}
The neutral currents $J_\mu(EM)$, $J_\mu(Z)$, $J_\mu(Z')$, and $J_\mu(Z'')$
associated with the Hamiltonian

\begin{eqnarray*}
H^0 &=& eA^\mu J_\mu(EM)+(g /{C_W})Z^\mu J_\mu(Z) \\
& &+ (g'/\sqrt{2})Z'^\mu J_\mu(Z') + (g/2)Z''^\mu J_\mu(Z''),
\end{eqnarray*}
\noindent
are

\begin{eqnarray}\nonumber
J_\mu(EM)&=&{2\over 3}\lbrace(\sum_{a=2}^3\bar{u}_a\gamma_\mu u_a)
+\bar{u}_1\gamma_\mu u_1+\bar{U}_1\gamma_\mu U_1 \\ \nonumber
& & +\bar{U'}_1\gamma_\mu U'_1\rbrace \\ \nonumber
& & -{1\over3}\lbrace(\sum_{a=2}^3\bar{d}_a\gamma_\mu d_a+ \bar{D}_a\gamma_\mu 
D_a+\bar{D'}_a\gamma_\mu D'_a) \\ \nonumber
& & +\bar{d}_1\gamma_\mu d_1\rbrace-\sum_{\alpha=1}^3\bar{e}^-_\alpha\gamma_\mu e^-_\alpha \\ \nonumber
&=&\sum_f q_f\bar{f}\gamma_\mu f,\\* \nonumber
J_\mu(Z)&=&J_{\mu,L}(Z)-S^2_WJ_\mu(EM),\\ \nonumber
J_\mu(Z')&=&T_WJ_\mu(EM)-J_{\mu,L}(Z'),\\ \nonumber
J_\mu(Z'')&=&\sum_{a=2}^3(\bar{D'}_{aL}\gamma_\mu 
D'_{aL}-\bar{D}_{aL}\gamma_\mu D_{aL})\\ \nonumber
& & +(-\bar{U'}_{1L}\gamma_\mu U'_{1L}+\bar{U}_{1L}\gamma_\mu U_{1L})\\ 
& & +\sum_{\alpha=1}^3(-\bar{N'}^0_{\alpha L}\gamma_\mu N'^0_{\alpha L}
+\bar{N}^0_{\alpha L}\gamma_\mu N^0_{\alpha L}),
\end{eqnarray}
\noindent where $e=gS_W=g'C_W\sqrt{1-T_W ^2/2}>0$ is the electric charge, 
$q_f$ is the electric charge of the fermion $f$ in units of $e$ and 
$J_\mu(EM)$ is the electromagnetic current. Notice that the $Z''_\mu$ 
current couples only to exotic fields. The left-handed currents are

\begin{eqnarray} \nonumber
J_{\mu,L}(Z)&=&{1\over 2}\lbrace \sum_{a=2}^3(\bar{u}_{aL}\gamma_\mu u_{aL}
-\bar{d}_{aL}\gamma_\mu d_{aL})\\ \nonumber
& & -(\bar{d}_{1L}\gamma_\mu d_{1L}- \bar{u}_{1L}\gamma_\mu u_{1L})\\ 
\nonumber
& & -\sum_{\alpha=1}^3(\bar{e}^-_{\alpha L}\gamma_\mu e^-_{\alpha L}
-\bar{\nu}_{e\alpha L}\gamma_\mu \nu_{e\alpha L})\rbrace \\ \nonumber 
&=&\sum_f T_{4f}\bar{f}_L\gamma_\mu f_L ,
\end{eqnarray}

\begin{eqnarray}\nonumber
J_{\mu,L}(Z')&=&(2T_{W})^{-1}\lbrace\sum_{a=2}^3\lbrack (1+T^2_W)\bar{u}_{aL}\gamma_\mu u_{aL} \\ \nonumber
& & +(1-T^2_W)\bar{d}_{aL}\gamma_\mu d_{aL} \\ \nonumber
& & -\bar{D}_{aL}\gamma_\mu D_{aL}-\bar{D'}_{aL}\gamma_\mu D'_{aL}\rbrack \\ \nonumber 
& & -(1+T^2_W)\bar{d}_{1L}\gamma_\mu d_{1L}-(1-T^2_W)\bar{u}_{1L}\gamma_\mu u_{1L} \\ \nonumber
& & +\bar{U}_{1L}\gamma_\mu U_{1L}
+\bar{U'}_{1L}\gamma_\mu U'_{1L}\\ \nonumber 
& & +\sum_{\alpha=1}^3\lbrack -(1+T^2_W)\bar{e}^-_{\alpha L}\gamma_\mu e^-_{\alpha L} \\ \nonumber
& & -(1-T^2_W)\bar{\nu}_{\alpha L}\gamma_\mu \nu_{\alpha L} \\ \nonumber
& & +\bar{N}^0_{\alpha L}\gamma_\mu N^0_{\alpha L} +\bar{N'}^0_{\alpha L}\gamma_\mu N'^0_{\alpha L}\rbrack \rbrace
\\ &=&\sum_f
T'_{4f}\bar{f}_L\gamma_\mu f_L,
\end{eqnarray}
where $T_{4f}=Dg(1/2,-1/2,0,0)$ is the third component of the weak isospin
and $T'_{4f}=(1/2T_W)Dg(1+T^2_W, 1-T^2_W, -1, -1)$= $T_W\lambda_3/2 +(1/T_W)(\lambda_8/\sqrt{3}+\lambda_{15}/\sqrt{6})$ is a convenient $4\times 4$ diagonal matrix, acting both of them on the representation 4 of $SU(4)_L$. Notice that $J_\mu(Z)$ is just the generalization of the neutral current present in the SM. This allows us to identify $Z_\mu$ as the neutral gauge boson of the SM, which is consistent with Eqs.~(\ref{fzzp}) and (\ref{y}).

The couplings of the mass eigenstates $Z_1^\mu$ and $Z_2^\mu$ are given by
\begin{eqnarray} \nonumber
H^{NC}&=&\frac{g}{2C_W}\sum_{i=1}^2Z_i^\mu\sum_f\{\bar{f}\gamma_\mu
[a_{iL}(f)(1-\gamma_5)\\ \nonumber & & 
+a_{iR}(f)(1+\gamma_5)]f\} \\ \nonumber
      &=&\frac{g}{2C_W}\sum_{i=1}^2Z_i^\mu\sum_f\{\bar{f}\gamma_\mu
      [g(f)_{iV}-g(f)_{iA}\gamma_5]f\},
\end{eqnarray}

\begin{table*}
\caption{\label{tab1}The $Z_1^\mu\longrightarrow \bar{f}f$ couplings.}
\begin{ruledtabular}
\begin{tabular}{lcc}
$f$ & $g(f)_{1V}$ & $g(f)_{1A}$ \\ \hline
$u_{2,3}$& $\cos\theta ({1\over 2}-{4S_W^2 \over 3})+
\frac{\sin\theta}{(3C_W^2-1)^{1/2}}({1\over 2} - {4 S_W^2\over 3})$
& ${1\over 2} \cos\theta + \sin\theta /[2(3C_W^2-1)^{1/2}]$ \\ 
$d_{2,3}$ & $(-{1\over
2}+{2S_W^2\over 3})\cos\theta+\frac{\sin\theta}{(3C_W^2-1)^{1/2}}(\frac{C^2_W}{2}+\frac{S^2_W}{6})$ 
& $-{1\over 2}\cos\theta + \frac{\sin\theta}{2(3C_W^2-1)^{1/2}}C_{2W}$ \\
$D_{2,3}$ & ${2S_W^2\over 3}\cos\theta-{1\over 2}\sin\theta (1 
- {7 S_W^2 \over 3})/(3C_W^2-1)^{1/2}$ &
$-{1\over 2}\sin\theta C_W^2/(3C_W^2-1)^{1/2}$ \\  
$D'_{2,3}$ & ${2S_W^2\over 3}\cos\theta-{1\over 2}\sin\theta (1 
- {7 S_W^2 \over 3})/(3C_W^2-1)^{1/2}$ &
$-{1\over 2}\sin\theta C_W^2/(3C_W^2-1)^{1/2}$ \\
$d_1$ & $(-{1\over
2}+{2S_W^2\over 3})\cos\theta-\sin\theta ({1 \over 2} 
- {2 S_W^2\over 3})/(3C_W^2-1)^{1/2}$ & $-{1\over 2}\cos\theta 
- \sin\theta /[2(3C_W^2-1)^{1/2}]$ \\
$u_1$& $\cos\theta ({1\over 2}-{4S_W^2 \over 3})-
\frac{\sin\theta}{(3C_W^2-1)^{1/2}}({C^2_W \over 2}+{5 S_W^2\over 6})$
& ${1\over 2} \cos\theta - \frac{\sin\theta}{2(3C_W^2-1)^{1/2}}C_{2W}$ \\ 
$U_1$& $-{4S_W^2\cos\theta
\over 3}+\sin\theta (1-{11\over 3}S_W^2)/[2(3C_W^2-1)^{1/2}]$ &
$C_W^2\sin\theta /[2(3C_W^2-1)^{1/2}]$ \\
$U'_1$& $-{4S_W^2\cos\theta
\over 3}+\sin\theta (1-{11\over 3}S_W^2)/[2(3C_W^2-1)^{1/2}]$ &
$C_W^2\sin\theta /[2(3C_W^2-1)^{1/2}]$ \\  
$e^-_{1,2,3}$& $\cos\theta
(-{1\over 2}+2S_W^2)- \frac{\sin\theta}{(3C_W^2-1)^{1/2}}({1\over 2} - 2S_W^2)$ & $ -{1\over 2}\cos\theta -\frac{\sin\theta}{2(3C_W^2-1)^{1/2}}$\\
$\nu_{1,2,3}$ &
${1\over 2}\cos\theta- \frac{\sin\theta}{2(3C_W^2-1)^{1/2}}C_{2W}$ &
$ {1\over 2}\cos\theta - \frac{\sin\theta}{2(3C_W^2-1)^{1/2}}C_{2W}$ \\ 
$N^0_{1,2,3}$ & $C_W^2\sin\theta /[2(3C_W^2-1)^{1/2}]$ &
$C_W^2\sin\theta /[2(3C_W^2-1)^{1/2}]$ \\ 
$N^{'0}_{1,2,3}$ &
$C_W^2\sin\theta /[2(3C_W^2-1)^{1/2}]$ &
$C_W^2\sin\theta /[2(3C_W^2-1)^{1/2}]$ \\ 
\end{tabular}
\end{ruledtabular}
\end{table*}

\begin{table*}
\caption{\label{tab2}The $Z_2^\mu\longrightarrow \bar{f}f$ couplings.}
\begin{ruledtabular}
\begin{tabular}{lcc}
$f$ & $g(f)_{2V}$ & $g(f)_{2A}$ \\ \hline
$u_{2,3}$& $-\sin\theta ({1\over 2}-{4S_W^2 \over 3})+
\frac{\cos\theta}{(3C_W^2-1)^{1/2}}({1\over 2} - {4 S_W^2\over 3})$
& $-{1\over 2} \sin\theta + \cos\theta /[2(3C_W^2-1)^{1/2}]$ \\
$d_{2,3}$ & $({1\over
2}-{2S_W^2\over 3})\sin\theta+\cos\theta /(3C_W^21)^{1/2} (\frac{C^2_W}{2}+\frac{S^2_W}{6})$ & 
${1\over 2}\sin\theta+ \frac{\cos\theta}{2(3C_W^2-1)^{1/2}}C_{2W}$ \\ 
$D_{2,3}$ & $-{2S_W^2\over 3}\sin\theta-{1\over 2}\cos\theta (1 - {7 S_W^2 
\over 3})/(3C_W^2-1)^{1/2}$ &
$-{1\over 2}\cos\theta C_W^2/(3C_W^2-1)^{1/2}$ \\
$D'_{2,3}$ & $-{2S_W^2\over 3}\sin\theta-{1\over 2}\cos\theta 
(1 - {7 S_W^2 \over 3})/(3C_W^2-1)^{1/2}$ &
$-{1\over 2}\cos\theta C_W^2/(3C_W^2-1)^{1/2}$ \\
$d_1$ & $({1\over
2}-{2S_W^2\over 3})\sin\theta-\cos\theta 
({1\over 2} - {2 S_W^2\over 3})/(3C_W^2-1)^{1/2}$ & 
${1\over 2}\sin\theta - \cos\theta /[2(3C_W^2-1)^{1/2}]$ \\
$u_1$& $-\sin\theta ({1\over 2}-{4S_W^2 \over 3})-
\frac{\cos\theta}{(3C_W^2-1)^{1/2}}({C^2_W \over 2}+{5 S_W^2\over 6})$
& $-{1\over 2} \sin\theta - \frac{\cos\theta}{2(3C_W^2-1)^{1/2}}C_{2W}$ \\  
$U_1$& ${4S_W^2\sin\theta
\over 3}+\cos\theta (1-{11\over 3}S_W^2)/[2(3C_W^2-1)^{1/2}]$ &
$C_W^2\cos\theta /[2(3C_W^2-1)^{1/2}]$ \\
$U'_1$& ${4S_W^2\sin\theta
\over 3}+\cos\theta (1-{11\over 3}S_W^2)/[2(3C_W^2-1)^{1/2}]$ &
$C_W^2\cos\theta /[2(3C_W^2-1)^{1/2}]$ \\
$e^-_{1,2,3}$& $\sin\theta
({1\over 2}-2S_W^2)- \frac{\cos\theta}{(3C_W^2-1)^{1/2}}({1\over 2} - 2S_W^2)$ & 
$ {1\over 2}\sin\theta-\frac{\cos\theta}{2(3C_W^2-1)^{1/2}}$\\ 
$\nu_{1,2,3}$ &
$-{1\over 2}\sin\theta -\frac{\cos\theta}{2(3C_W^2-1)^{1/2}}C_{2W}$ &
$-{1\over 2}\sin\theta -\frac{\cos\theta}{2(3C_W^2-1)^{1/2}}C_{2W}$ \\ 
$N^0_{1,2,3}$ & $C_W^2\cos\theta /[2(3C_W^2-1)^{1/2}]$ &
$C_W^2\cos\theta /[2(3C_W^2-1)^{1/2}]$ \\ 
$N^{'0}_{1,2,3}$ &
$C_W^2\cos\theta /[2(3C_W^2-1)^{1/2}]$ &
$C_W^2\cos\theta /[2(3C_W^2-1)^{1/2}]$ \\ 
\end{tabular}
\end{ruledtabular}
\end{table*}
\noindent
where
\begin{eqnarray} \nonumber
a_{1L}(f)&=&\cos\theta(T_{4f}-q_fS^2_W)\\ \nonumber & &
+\frac{g'\sin\theta
C_W}{g\sqrt{2}} (T'_{4f}-q_fT_W)\;, \\ \nonumber
a_{1R}(f)&=&-q_fS_W\left(\cos\theta
S_W+\frac{g'\sin\theta}{g\sqrt{2}}\right)\;,\\ \nonumber
a_{2L}(f)&=&-\sin\theta(T_{4f}-q_fS^2_W)\\ \nonumber & &
+\frac{g'\cos\theta
C_W}{g\sqrt{2}} (T'_{4f}-q_fT_W)\;, \\ \label{a}
a_{2R}(f)&=&q_fS_W\left(\sin\theta S_W-\frac{g'\cos\theta}{g\sqrt{2}}\right),
\end{eqnarray}
and
\begin{eqnarray} \nonumber
g(f)_{1V}&=&\cos\theta(T_{4f}-2S_W^2q_f)\\ \nonumber & &
+\frac{g'\sin\theta}{g\sqrt{2}}
(T'_{4f}C_W-2q_fS_W)\;, \\ \nonumber
g(f)_{2V}&=&-\sin\theta(T_{4f}-2S_W^2q_f)\\ \nonumber & &
+\frac{g'\cos\theta}{g\sqrt{2}}
(T'_{4f}C_W-2q_fS_W) \;,\\ \nonumber g(f)_{1A}&=&\cos\theta
T_{4f}+\frac{g'\sin\theta}{g\sqrt{2}}T'_{4f}C_W\;, \\ \label{g}
g(f)_{2A}&=&-\sin\theta
T_{4f}+\frac{g'\cos\theta}{g\sqrt{2}}T'_{4f}C_W.
\end{eqnarray}
The values of $g_{iV},\; g_{iA}$ with $i=1,2$ are listed in Tables \ref{tab1} and \ref{tab2}.

As we can see, in the limit $\theta=0$ the couplings of $Z_1^\mu$ to 
the ordinary leptons and quarks are the same as in the SM; due to 
this we can test the new physics beyond the SM predicted by this 
particular model.


\section{\label{sec:sec5}Fermion masses}
The Higgs scalars introduced in Sec.~\ref{sec:sec3} not only break the 
symmetry in an appropriate way, but produce the following mass terms for 
the fermions of the model.
\subsection{Quark masses}
For the quark sector we can write the following Yukawa terms:
\begin{eqnarray} \nonumber
{\cal L}^Q_Y&=& \sum_{a=2}^3Q^T_{aL}C\{\phi^*_1 (\sum_{\alpha=1}^3 
h^a_{u\alpha}u^c_{\alpha L} + h^a_UU^c_L + h^{a\prime}_UU_L^{\prime c}) 
\\ \nonumber 
&+& 
(\phi_2+\phi_3)[\sum_{\alpha=1}^3h^a_{\alpha d} d^c_{\alpha L} + 
\sum_{b=2}^3(h^a_{bD}D^c_{bL} + h_{bD}^{a\prime}D_{bL}^{\prime c})]\}
\\ \nonumber 
&+& Q^T_{1L}C\{\phi_1[\sum_{\alpha=1}^3 h^1_{d\alpha}d^c_{\alpha L} + 
\sum_{a=2}^3(h^1_{aD}D^c_{aL} + h_{aD}^{1\prime}D_{aL}^{\prime c})]
\\ \nonumber
&+& (\phi_2^* + \phi_3^*)
(\sum_{\alpha=1}^3 
h^1_{u\alpha}u^c_{\alpha L} + h^1_UU^c_L + h^{1\prime}_UU_L^{\prime c})\} 
+ h.c., 
\end{eqnarray}
where the $h's$ are Yukawa couplings and $C$ is the charge conjugate 
operator. This Lagrangian produces the following tree level quark masses:
\begin{itemize}
\item $U_1^\prime, D_2^\prime$ and $D_3^\prime$ acquire heavy masses of 
the order of $V^\prime >>v$.
\item $U_1, D_2$ and $D_3$ acquire heavy masses of the order of $V>>v$.
\item $u_3, u_2$ and $d_1$ acquire masses of the order of $v\approx 174$ GeV.
\item $u_1, d_2$ and $d_3$ remain massless at the tree level.
\end{itemize}
The former mass spectrum is far from being realistic, but it can be 
improved by implementing the following program:
\begin{enumerate}
\item To introduce a discrete symmetry in order to avoid a tree-level 
mass for $d_1$ (and maybe for $u_2$ too).
\item To introduce a new Higgs field $\phi_4[1,\bar{4},-1/4]$ which does  
not acquire VEV but that introduces a quartic coupling 
$\phi_1^*\phi_2\phi_3\phi_4$ in the Higgs potential in order to generate 
radiative masses for the ordinary quarks.
\item To tune the Yukawa couplings in order to obtain the correct mixing 
between flavors (ordinary and exotic) with the same electric charge.
\end{enumerate}

\subsection{Lepton masses}
For the charged leptons we have the following Yukawa terms:
\begin{equation}
{\cal L}_Y^l=\sum_{\alpha=1}^3\sum_{\beta=1}^3h_{\alpha\beta}^e 
L_{\alpha L}^TC\phi_1 e_{\beta L}^+ + h.c..
\end{equation}
Notice that for $h^e_{\alpha\beta}=h\delta_{\alpha\beta}$ we get a mass only for the heaviest lepton (the $\tau$). So, in the context of this model the masses for the charged leptons can be generated in a consistent way, with the masses for $e^-$ and $\mu^-$ suppressed by differences of Yukawa couplings. 

The neutral leptons remain massless as far as we use only the original
fields introduced in Sec.~\ref{sec:sec2}. But as mentioned earlier, we may
introduce one or more Weyl singlet states $N^0_{L,b},\; b=1,2,...$ which may implement the appropriate neutrino oscillations \cite{neutrinos}.

 
\section{\label{sec:sec6}
Constrains on the $(Z^\mu-Z'^{\mu})$ mixing angle and the $Z^{\mu}_2$ 
mass} 
To bound $\sin\theta$ and $M_{Z_2}$ we use parameters
measured at the $Z$ pole from CERN $e^+e^-$ collider (LEP), SLAC Linear 
Collider (SLC), and atomic parity violation constraints which are given
in Table \ref{tab3}. 

The expression for the partial decay width for $Z^{\mu}_1\rightarrow
f\bar{f}$ is
 
\begin{eqnarray}\nonumber
\Gamma(Z^{\mu}_1\rightarrow f\bar{f})&=&\frac{N_C G_F
M_{Z_1}^3}{6\pi\sqrt{2}}\rho \Big\{\frac{3\beta-\beta^3}{2}
[g(f)_{1V}]^2 \\ \label{ancho}
& + & \beta^3[g(f)_{1A}]^2 \Big\}(1+\delta_f)R_{EW}R_{QCD}, \quad
\end{eqnarray}

\noindent 
where $f$ is an ordinary SM fermion, $Z^\mu_1$ is the physical gauge boson
observed at LEP, $N_C=1$ for leptons while for quarks
$N_C=3(1+\alpha_s/\pi + 1.405\alpha_s^2/\pi^2 - 12.77\alpha_s^3/\pi^3)$,
where the 3 is due to color and the factor in parenthesis represents the
universal part of the QCD corrections for massless quarks 
(for fermion mass effects and further QCD corrections which are 
different for vector and axial-vector partial widths, see 
Ref.~\cite{kuhn}); $R_{EW}$ are the electroweak corrections which include the leading order QED corrections given by $R_{QED}=1+3\alpha/(4\pi)$. $R_{QCD}$ are further QCD corrections (for a comprehensive review see Ref.~\cite{leike} and references therein), and $\beta=\sqrt{1-4 m_f^2/M_{Z_1}^2}$ is a kinematic factor which can be taken equal to $1$ for all the SM fermions except for the bottom quark. 
The factor $\delta_f$ contains the one loop vertex
contribution which is negligible for all fermion fields except for the 
bottom quark for which the contribution coming from the top quark at the 
one loop vertex radiative correction is parametrized as $\delta_b\approx 
10^{-2} [-m_t^2/(2 M_{Z_1}^2)+1/5]$ \cite{pich}. The $\rho$ parameter 
can be expanded as $\rho = 1+\delta\rho_0 + \delta\rho_V$ where the 
oblique correction $\delta\rho_0$ is given by
$\delta\rho_0\approx 3G_F m_t^2/(8\pi^2\sqrt{2})$, and $\delta\rho_V$ is 
the tree level contribution due to the $(Z_{\mu} - Z'_{\mu})$ mixing which 
can be parametrized as $\delta\rho_V\approx
(M_{Z_2}^2/M_{Z_1}^2-1)\sin^2\theta$. Finally, $g(f)_{1V}$ and $g(f)_{1A}$
are the coupling constants of the physical $Z_1^\mu$ field with ordinary
fermions which are listed in Table \ref{tab1}.

In what follows we are going to use the experimental values \cite{pdg}:
$M_{Z_1}=91.188$ GeV, $m_t=174.3$ GeV, $\alpha_s(m_Z)=0.1192$, 
$\alpha(m_Z)^{-1}=127.938$, and
$\sin^2\theta_W=0.2333$. The experimental values are introduced using the
definitions $R_\eta\equiv \Gamma(\eta\eta)/\Gamma(hadrons) $ for
$\eta=e,\mu,\tau,b,c$.

As a first result notice from Table \ref{tab1}, that our model predicts 
$R_e=R_\mu=R_\tau$, in agreement with the experimental results in Table 
\ref{tab3}.

The effective weak charge in atomic parity violation, $Q_W$, can be 
expressed as a function of the number of protons $(Z)$ and the number of 
neutrons $(N)$ in the atomic nucleus in the form 

\begin{equation}
Q_W=-2\left[(2Z+N)c_{1u}+(Z+2N)c_{1d}\right], 
\end{equation}
\noindent
where $c_{1q}=2g(e)_{1A}g(q)_{1V}$. The theoretical value for $Q_W$ for 
the Cesium atom is given by \cite{sirlin} $Q_W(^{133}_{55}Cs)=-73.09\pm0.04 
+ \Delta Q_W$, where the contribution of new physics is included in $\Delta 
Q_W$ which can be written as \cite{durkin}

\begin{equation}\label{DQ} 
\Delta 
Q_W=\left[\left(1+4\frac{S^4_W}{1-2S^2_W}\right)Z-N\right]\delta\rho_V
+\Delta Q^\prime_W.
\end{equation}

The term $\Delta Q^\prime_W$ is model dependent and it can be obtained 
for our model by using $g(e)_{iA}$ and $g(q)_{iV}$, $i=1,2$ from Tables \ref{tab1} and \ref{tab2}. The value we obtain is

\begin{equation}
\Delta Q_W^\prime=(10.29 Z + 12.40 N) \sin\theta + (9.23 Z + 7.79 N)
\frac{M^2_{Z_1}}{M^2_{Z_2}}\; .
\end{equation}

The discrepancy between the SM and the experimental data for $\Delta Q_W$ 
is given by \cite{casal}

\begin{equation}
\Delta Q_W=Q^{exp}_W-Q^{SM}_W=1.03\pm 0.44,
\end{equation}
which is $2.3\; \sigma$ away from the SM predictions.

\begin{table}
\caption{\label{tab3}Experimental data and SM values for the parameters.}
\begin{ruledtabular}
\begin{tabular}{lcl}
& Experimental results & SM \\ \hline
$\Gamma_Z$(GeV)  & $2.4952 \pm 0.0023$  &  $2.4966 \pm 0.0016$  \\   
$\Gamma(had)$ (GeV)  & $1.7444 \pm 0.0020$ & $1.7429 \pm 0.0015$ \\ 
$\Gamma(l^+l^-)$ (MeV) & $83.984\pm 0.086$ & $84.019 \pm 0.027$ \\
$R_e$ & $20.804\pm 0.050$ & $20.744\pm 0.018$ \\ 
$R_\mu$ & $20.785\pm 0.033$ & $20.744\pm 0.018$ \\ 
$R_\tau$ & $20.764\pm 0.045$ & $20.790\pm 0.018$ \\ 
$R_b$ & $0.21664\pm 0.00068$ & $0.21569\pm 0.00016$ \\ 
$R_c$ & $0.1729\pm 0.0032$ & $0.17230\pm 0.00007$ \\ 
$Q_W^{Cs}$ & $-72.65\pm 0.28\pm 0.34$ & $-73.10\pm 0.03$ \\
$M_{Z_{1}}$(GeV) & $ 91.1872 \pm 0.0021 $ & $ 91.1870 \pm 0.0021 $ \\
\end{tabular}
\end{ruledtabular}
\end{table}

\begin{figure} 
\begin{center}
\includegraphics{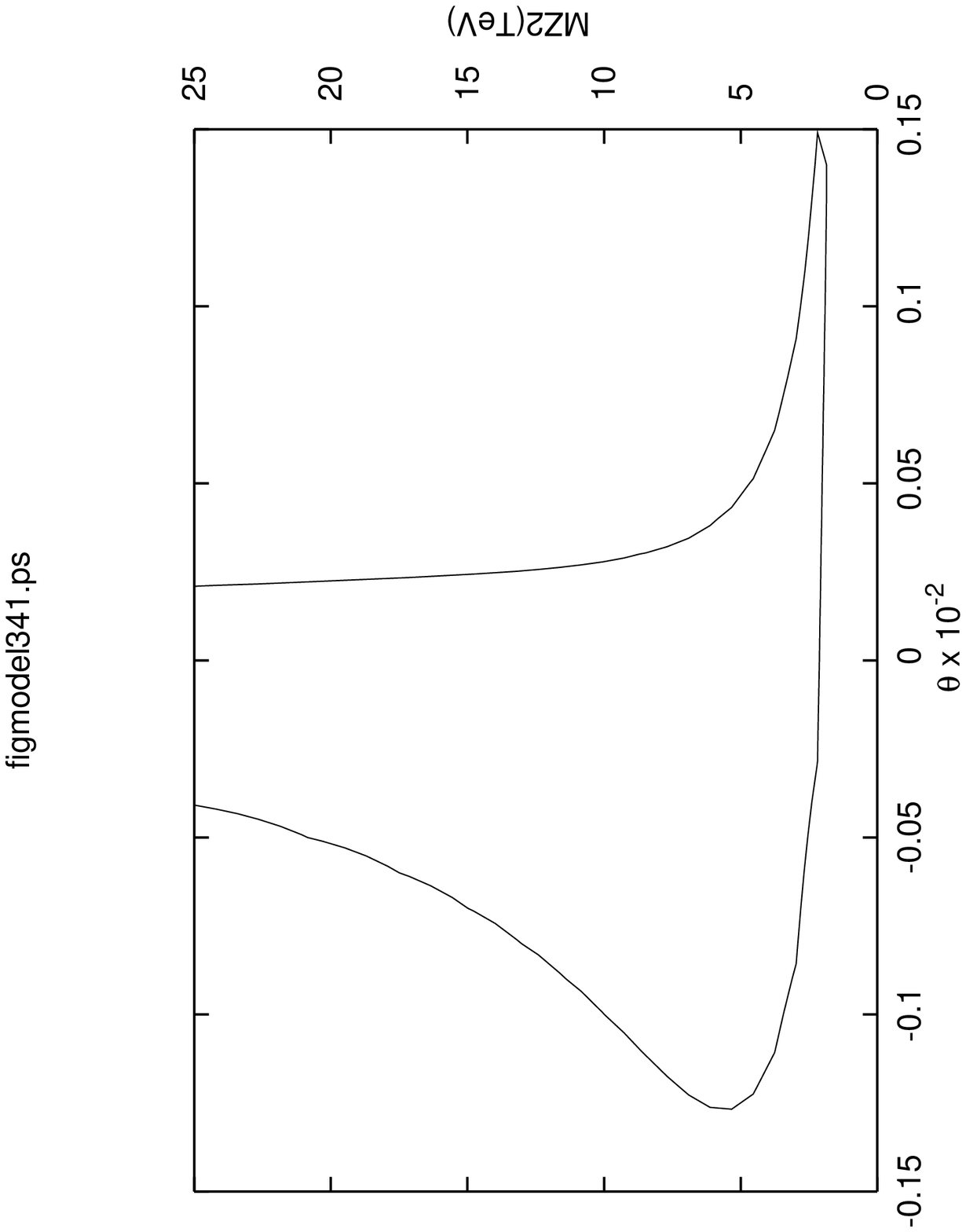}
\caption{\label{fig1} Contour plot displaying the allowed region for
$\theta$ vs. $M_{Z_2}$ at $95\%$ C.L.}
\end{center}
\end{figure}

Introducing the expressions for $Z$ pole observables in Eq.~(\ref{ancho}), 
with $\Delta Q_W$ in terms of new physics in Eq.~(\ref{DQ}) and using 
experimental data from LEP, SLC and atomic parity violation (see Table 
\ref{tab3}), we do a $\chi^2$ fit and we find the best allowed region in the $(\theta-M_{Z_2})$ plane at $95\%$ confidence level (C.L.). In 
Fig.~\ref{fig1} we display this region which gives us the constraints 
\begin{equation} 
-0.0013\leq\theta\leq 0.0015, \;\;\; 1.85\; {\mbox TeV} \leq M_{Z_2}.
\end{equation} 
As we can see the mass of the new neutral gauge boson is compatible 
with the bound obtained in $p\bar{p}$ collisions at the Fermilab 
Tevatron \cite{abe}. From our analysis we can also see that for $\vert \theta \vert \rightarrow 0$, $M_{Z_2}$ peaks at a finite value larger than $100$ TeV which still copes with experimental constraints on the $\rho$ parameter.
 

\section{Conclusions}
We have presented an anomaly-free model based on the local gauge
group $SU(3)_c\otimes SU(4)_L\otimes U(1)_X$. We break the
gauge symmetry down to $SU(3)_c\otimes U(1)_{Q}$ and at the same
time give masses to the fermion fields in the model in a
consistent way by using three different Higgs scalars $\phi_i,\:
i=1,2,3,$ which set two different mass scales: $V\sim V^\prime >> v =174$
GeV. By using experimental results we bound the mixing angle $\theta$ 
between the SM neutral current and a new one 
to be $-0.0013<\theta<0.0015$ and the lowest bound
for $M_{Z_2}$ is $1.85$ TeV $\leq M_{Z_2}$.

Our model includes four exotic down type quarks $D_a, D_a^\prime, \;\;  
a=2,3$ of electric charge $-1/3$ and two exotic up quarks
$U_1,U_1^\prime$ of electric charge 2/3. The six exotic quarks acquire
large masses of the order of $V\simeq V^\prime >>v=174$ GeV and are usefull in two ways: first, they mix with the ordinary up and down quarks in the three families with a mixing that can be used in order to produce a consistent mass spectrum (masses and mixings)  for ordinary quarks; second, they can be used in order to implement the so-called little Higgs
mechanism \cite{little}.

Notice also the consistence of our model in the charged lepton sector. Not only it predicts the correct ratios $R_\eta,\;\;\eta=e,\mu,\tau,$ in the $Z$ decays, but the model also allows for a consistent mass pattern of the particles, which do not include leptons with exotic electric charges.

In the main body of this paper we have studied an specific model based on 
the 3-4-1 gauge structure. This model is just one of a large variety of 
models based on the same gauge group. A systematic analysis of 
models without exotic electric charges with the same gauge structure is 
presented in the Appendix at the end of the paper. A phenomenological 
analysis for all those model can be done, but we think it is not profitable since all of them must produce similar results at low energies.

\section*{ACKNOWLEDGMENTS}
We thank Jorge I. Zuluaga for helping us with the numerical analysis 
presented in Sec.\ref{sec:sec6}. W.A.P. thanks the Theoretical Physics 
Laboratory at the Universidad de La Plata in La Plata, Argentina, where 
part of this work was done.

\appendix*
\section{}
In what follows we present a systematic analysis of  
models without exotic electric charges, based on the local gauge structure 
$SU(3)_c\otimes SU(4)_L\otimes U(1)_X$. 

We assume that the electroweak group is 
$SU(4)_L\otimes U(1)_X\supset SU(3)_L \otimes U(1)_Z 
\supset SU(2)_L \otimes U(1)_Y$, where the gauge structure 
$SU(3)_L\otimes U(1)_Z$ refers to the one presented in Ref.~\cite{331}. 
We also assume that the left handed quarks (color triplets), left-handed 
leptons (color singlets) and scalars, transform either under the 4 or 
the $\bar{4}$ fundamental representations of $SU(4)_L$. Two classes of 
models will be discussed: one family models where the anomalies cancel in 
each family as in the SM, and family models where the anomalies cancel 
by an interplay between the several families. As in the SM, $SU(3)_c$ is 
vectorlike.

The most general expression for the electric charge generator in
$SU(4)_L\otimes U(1)_X$ is a linear combination of the four diagonal
generators of the gauge group
\begin{equation}\label{ch}
Q=aT_{3L}+\frac{1}{\sqrt{3}}bT_{8L}+ \frac{1}{\sqrt{6}}cT_{15L}+ XI_4, 
\end{equation} 
where $T_{iL}=\lambda_{iL}/2$, being $\lambda_{iL}$ the Gell-Mann matrices
for $SU(4)_L$ normalized as  Tr$(\lambda_i\lambda_j)=2\delta_{ij}$,
$I_4=Dg(1,1,1,1)$ is the diagonal $4\times 4$ unit matrix, and $a$, $b$ 
and $c$ are free parameters to be fixed next. Notice that we can absorb  
an eventual coefficient for $X$ in its definition.

If we assume that the usual isospin $SU(2)_L$ of the SM is such that
$SU(2)_L\subset SU(4)_L$, then $a=1$ and we have just a two-parameter set
of models, all of them characterized by the values of $b$ and $c$. So, 
Eq.~(\ref{ch}) allows for an infinite number of models in the context of the 3-4-1 theory, each one associated to particular values of the parameters $b$ and $c$, with characteristic signatures that make them different from each other. 

There are a total of 24 gauge bosons in the gauge group under
consideration, 15 of them associated with $SU(4)_L$ which can be written as:
\begin{widetext}
\begin{equation}\nonumber
{1\over 2}\lambda_\alpha A^\alpha_\mu ={1\over \sqrt{2}} 
\left(\begin{array}{cccc}D^0_{1\mu} & W^+_\mu & K^{(b+1)/2}_\mu & X^{(3+b
+2c)/6}_\mu\\ W^-_\mu &
D^0_{2\mu} &  K^{(b-1)/2}_\mu &  X^{(-3+b+2c)/6}_\mu \\
K^{-(b+1)/2}_\mu & \bar{K}^{-(b-1)/2}_\mu & D^0_{3\mu} & Y^{-(b-c)/3}_\mu\\
X^{-(3+b+2c)/6}_\mu & \bar{X}^{(3-b-2c)/6}_\mu & \bar{Y}^{(b-c)/3}_\mu & 
D^0_{4\mu} \end{array}\right),  \nonumber
\end{equation}
\end{widetext}
where $D^\mu_1=A_3^\mu/\sqrt{2}+A_8^\mu/\sqrt{6}+A_{15}^\mu/\sqrt{12},\;
D^\mu_2=-A_3^\mu/\sqrt{2}+A_8^\mu/\sqrt{6}+A_{15}^\mu/\sqrt{12}$;
$D^\mu_3=-2A_8^\mu/\sqrt{6}+A_{15}^\mu/\sqrt{12}$, and 
$D^\mu_4=-3 A_{15}^\mu/\sqrt{12}$. The upper indices in the gauge bosons in
the former expression stand for the electric charge of the corresponding 
particle, some of them functions of the $b$ and $c$ parameters as they 
should be. Notice that if we demand for gauge bosons with electric 
charges $0, \pm 1$ only, there are not more than four different possibilities 
for the simultaneous values of $b$ and $c$; they are: 
$b= c = 1$; $b = c = -1$; $b = 1$, $c = -2$, and $b = -1$, $c = 2$.

Now, contrary to the SM where only the abelian $U(1)_Y$ factor is
anomalous, in the 3-4-1 theory both, $SU(4)_L$ and $U(1)_X$ are anomalous
($SU(3)_c$ is vectorlike). So, special combinations of multiplets must be
used in each particular model in order to cancel the possible
anomalies, and obtain renormalizable models. The triangle anomalies
we must take care of are: $[SU(4)_L]^3$, $[SU(3)_c]^2U(1)_X$, 
$[SU(4)_L]^2U(1)_X$, $[grav]^2U(1)_X$ and $[U(1)_X]^3$.

Now let us see how the charge operator in Eq.~(\ref{ch}) acts on the
representations 4 and $\bar{4}$ of $SU(4)_L$:

\begin{eqnarray*}
Q[4]
&=&Dg.({1\over 2}+{b\over 6}+{c\over 12}+X, -{1\over 2}+{b\over 6}+
{c\over 12}+X, \\
& & -{2b\over 6}+{c\over 12}+X, -{3c \over 12}+X),\\
Q[\bar{4}]
&=&Dg.(-{1\over 2}-{b\over 6}-{c\over 12}+X, {1\over 2}-{b\over 6}-
{c\over 12}+X, \\
& & {2b\over 6}-{c\over 12}+X, {3c \over 12}+X).
\end{eqnarray*}

\noindent 
Notice that, if we accommodate the known
left-handed quark and lepton isodoublets in the two upper components of 4
and $\bar{4}$ (or $\bar{4}$ and 4), do not allow for electrically charged
antiparticles in the two lower components of the multiplets (antiquarks
violate $SU(3)_c$ and $e^+, \mu^+$ and $\tau^+$ violate lepton number at
the tree level) and forbid the presence of exotic electric charges in the
possible models, then the electric charge of the third and fourth
components in $4$ and $\bar{4}$ must be equal either to the charge of the
first and/or second component, which in turn implies that $b$ and $c$ can
take only the four sets of values stated above. So, these four sets of
values for $b$ and $c$ are necessary and sufficient conditions in order to
exclude exotic electric charges in the fermion sector too.

A further analysis also shows that models with $b=c=-1$ are equivalent,
via charge conjugation, to models with $b=c=1$. Similarly, models 
with $b=-1,\; c=2$ are equivalent to models with $b=1,\; c=-2$. So, 
with the constraints impossed, we have only two different sets of models; 
those for $b=c=1$ and those for $b=1, \; c=-2$.

\begin{table*}
\caption{\label{tab4}Anomalies for sets with values $b$=$c$=1}
\begin{ruledtabular}
\begin{tabular}{lcccccc} 
Anomaly & $S_1^q$ & $S_2^q$ & $S_3^l$ & $S_4^l$ & $S_5^l$ & $S_6^l$ \\ 
\hline
$[U(1)_X]^3$ & $-9/16$ & $-27/16$ & 21/16 & $-15/16$ & 15/16 & $-21/16$ \\
$[SU(4)_L]^2U(1)_X$ & $-1/4$ & 5/4 & $-3/4$ & 1/4 & $-1/4$ & 3/4 \\
$[SU(4)_L]^3 $ & 3 & $-3$ & 1 & 1 & $-1$ & $-1$ \\  
\end{tabular}
\end{ruledtabular}
\end{table*}

\begin{table*}
\caption{\label{tab5}Anomalies for sets with values $b=1, \; c=-2$}
\begin{ruledtabular}
\begin{tabular}{lcccccc} 
Anomaly & $S_1^q$ & $S_2^q$ & $S_3^l$ & $S_4^l$ & $S_5^l$ & $S_6^l$ \\ 
\hline
$[U(1)_X]^3$ & $-3/2$ & $-3/2$ & 3/2 & $ 3/2 $ & -3/2 & $-3/2 $ \\
$[SU(4)_L]^2U(1)_X$ & 1/2 & 1/2 & $-1/2$ & $-1/2$ & 1/2 & 1/2 \\
$[SU(4)_L]^3 $ & 3 & $-3$ & 1 & $-1$ & 1 & $-1$ \\  
\end{tabular}
\end{ruledtabular}
\end{table*}

\subsection{Models for $b=c=1$}
First let us define the following complete sets of spin 1/2 Weyl spinors 
(complete in the sense that each set contains its own charged 
antiparticles):
\begin{itemize}
\item $S_1^q=\{(u,d,D,D^\prime)_L\sim [3,4,-\frac{1}{12}], \; 
u_L^c \sim [\bar{3}, 1, -{2\over 3}], \; 
d_L^c \sim [\bar{3},1,{1\over 3}], \; 
D_L^c \sim [\bar{3}, 1, {1\over 3}], \; 
D_L^{\prime c} \sim [\bar{3}, 1, {1\over 3}]\}.$ 
\item $S_2^q=\{(d,u,U,U^\prime)_L\sim [3,\bar{4},\frac{5}{12}], \; 
u_L^c \sim [\bar{3}, 1, -{2\over 3}], \; 
d_L^c \sim [\bar{3},1,{1\over 3}], \; 
U_L^c \sim [\bar{3}, 1, -{2\over 3}], \; 
U_L^{\prime c} \sim [\bar{3},1,-{2\over 3}]\}.$ 
\item $S_3^l=\{ (\nu^0_e, e^-, E^-, E^{\prime -})_L \sim 
[1,4,-{3 \over 4}],\;
e^+_L \sim [1,1,1], \; E^+_L \sim [1,1,1], \; E^{\prime +}_L \sim 
[1,1,1]\}.$ 
\item $S_4^l = \{ (E^+, N^0_1, N^0_2, N^0_3)_L\sim 
[1, 4, {1\over 4}], \; E^-_L\sim [1,1,-1]\}.$ 
\item $S_5^l = \{ (e^-, \nu^0_e, N^0, N^{\prime 0})_L\sim 
[1, \bar{4}, -{1\over 4}], \; e^+_L\sim [1,1,1]\}.$ 
\item $S_6^l=\{ (N, E_1^+, E_2^+, E_3^+)_L \sim [1, \bar{4}, 
{3 \over 4}],\;
E_{1L}^- \sim [1,1,-1], \; E_{2L}^- \sim [1,1,-1], \; E_{3L}^- \sim 
[1,1,-1]\}.$
\end{itemize}

Due to the fact that each set includes charged particles together with
their corresponding antiparticles, and since $SU(3)_c$ is vectorlike, the
anomalies $[grav]^2U(1)_X, \; [SU(3)_c]^3$ and $[SU(3)_c]^2U(1)_X$
automatically vanish. So, we only have to take care of the remaining three
anomalies whose values are shown in Table \ref{tab4}.

Several anomaly free models can be constructed from this table. Let us 
see.
\subsubsection{Three family models}
We found two three family structures which are:
\begin{itemize}
\item Model {\bf A} = $2S_1^q \oplus S_2^q \oplus 3S_5^l$. (The model 
analyzed in the main text.) 
\item Model {\bf B} = $S_1^q \oplus 2S_2^q \oplus 3S_3^l$. 
\end{itemize}
\subsubsection{Two family models}
We find only one two family structure given by:
Model {\bf C} = $S_1^q \oplus S_2^q \oplus S_3^l \oplus S_5^l$. 
\subsubsection{One family models}
A one family model can not be directly extracted from $S_i, \; i=1,2,...,6$, but we can check that the following particular arrangement is an anomaly free one family structure:
Model {\bf D}=$S_1^q\oplus (e^-,\nu_e^0, N^0, N^{\prime 0})_L \oplus 
(E_1^-, N_1^0, N_2^0, N_3^0)_L \oplus (N_4^0, E_1^+, e^+, E_2^+)_L \oplus E_2^-.$ As it can be checked, this model reduces to the model in 
Ref.~\cite{spm} for the breaking chain $SU(4)_L\otimes U(1)_X\longrightarrow SU(3)_L\otimes U(1)_\alpha\otimes U(1)_X\longrightarrow SU(3)_L\otimes U(1)_Z$, for the value $\alpha = 1/12$. In an analogous way, other one family models with more exotic charged leptons can also be constructed.

\subsection{Models for $b=1, c=-2$}
As in the previous case, let us define the following complete sets of spin 
1/2 Weyl spinors:
\begin{itemize}
\item $S_1^{\prime q}=\{(u,d,D,U)_L\sim [3,4,\frac{1}{6}], \; 
u_L^c \sim [\bar{3}, 1, -{2\over 3}], \; 
d_L^c \sim [\bar{3},1,{1\over 3}], \; 
D_L^c \sim [\bar{3}, 1, {1\over 3}], \; 
U_L^{c} \sim [\bar{3}, 1, -{2\over 3}]\}.$ 
\item $S_2^{\prime q}=\{(d,u,U,D)_L\sim [3,\bar{4},\frac{1}{6}], \; 
u_L^c \sim [\bar{3}, 1, -{2\over 3}], \; 
d_L^c \sim [\bar{3},1,{1\over 3}], \; 
U_L^c \sim [\bar{3}, 1, -{2\over 3}], \; 
D_L^{c} \sim [\bar{3},1,{1\over 3}]\}.$ 
\item $S_3^{\prime l}=\{ (\nu^0_e, e^-, E^-, N^0)_L \sim 
[1,4,-{1 \over 2}],\;
e^+_L \sim [1,1,1], \; E^+_L \sim [1,1,1]\}.$ 
\item $S_4^{\prime l} = \{ (e^-, \nu^0_e, N^0, E^-)_L\sim 
[1, \bar{4}, -{1\over 2}], \; e^+_L\sim [1,1,1],\; E^+_L\sim [1,1,1]\}.$ 
\item $S_5^{\prime l} = \{(E^+, N^0_1, N^0_2, e^+)_L\sim 
[1, 4, {1\over 2}], \; E^-_L\sim [1,1,-1],\; e^-_L\sim [1,1,-1]\}.$ 
\item $S_6^{\prime l}=\{ (N_3^0, E^+, e^+, N_4^0)_L \sim 
[1, \bar{4}, {1 \over 2}],\;
E_L^- \sim [1,1,-1], \; e_{L}^- \sim [1,1,-1]\}.$
\end{itemize}

For the former sets the anomalies $[grav]^2U(1)_X, \; [SU(3)_c]^3$ and
$[SU(3)_c]^2U(1)_X$ vanish.  The other anomalies are shown in Table 
\ref{tab5}. Again, several anomaly free models can be constructed from 
this table. Let us see.

\subsubsection{Three family models}
We found two three family structures which are:

\begin{itemize}
\item Model {\bf E} = $2S_1^{\prime q} \oplus S_2^{\prime q} \oplus 
3S_4^{\prime l}$. 
\item Model {\bf F} = $S_1^{\prime q} \oplus 2S_2^{\prime q} \oplus 
3S_3^{\prime l}$. 
\end{itemize}

\subsubsection{Two family models}
We find again only one two family structure given by:
Model {\bf G} = $S_1^{\prime q} \oplus S_2^{\prime q}\oplus 
S_3^{\prime l}\oplus S_4^{\prime l}$. 
\subsubsection{One family models}
Two one family models can be constructed using $S^\prime_i, \; i=1,....,6$. 
They are:
\begin{itemize}
\item Model {\bf H}= $S^{\prime q}_2\oplus 2S^{\prime l}_3\oplus S^{\prime l}_5$.
\item Model {\bf I}= $S^{\prime q}_1\oplus 2S^{\prime l}_4\oplus S^{\prime l}_6$.
\end{itemize}
To conclude this appendix let us mention that for the values of the 
parameters $b$ and $c$ used in our analysis, many more anomaly-free 
models  can be constructed, all of them featuring the SM phenomenology at 
energies below 100 GeV. Model {\bf A}, discussed in the main text, is 
just one example.

\end{document}